\documentclass[conference, a4paper]{IEEEtran}
\IEEEoverridecommandlockouts
\usepackage[utf8]{inputenc}
\usepackage{color}
\usepackage{xcolor}
\usepackage{array}
\usepackage{verbatim}
\usepackage{float}
\usepackage{amsmath}
\usepackage{amsthm}
\usepackage{amssymb}
\usepackage{graphicx}
\usepackage{longtable}
\usepackage{multirow}
\usepackage{booktabs}
\usepackage[unicode=true,
bookmarks=false,
breaklinks=false,pdfborder={0 0 1},colorlinks=false]
{hyperref}
\hypersetup{
	colorlinks,bookmarksopen,bookmarksnumbered,citecolor=blue,urlcolor=blue}
\usepackage{cite}

\usepackage{array}
\usepackage{overpic}
\usepackage{dblfloatfix}
\usepackage{balance}

\usepackage{lipsum}
\usepackage{mathtools}
\usepackage{cuted}

\usepackage{algorithmic}
\usepackage{longtable}

\floatstyle{ruled}
\newfloat{algorithm}{tbp}{loa}
\providecommand{\algorithmname}{Algorithm}
\floatname{algorithm}{\protect\algorithmname}

\makeatletter
\let\oldforeign@language\foreign@language
\DeclareRobustCommand{\foreign@language}[1]{%
	\lowercase{\oldforeign@language{#1}}}

\let\oldforeign@language\foreign@language
\DeclareRobustCommand{\foreign@language}[1]{%
	\lowercase{\oldforeign@language{#1}}}

\ifCLASSINFOpdf
\else
\fi

\@ifundefined{showcaptionsetup}{}{%
	\PassOptionsToPackage{caption=false}{subfig}}
\usepackage{subfig}

\usepackage{balance}

\ifCLASSINFOpdf
\else
\fi

\ifCLASSINFOpdf
\else
\fi

\def\ps@IEEEtitlepagestyle{%
	\def\@oddhead{\parbox[t][\height][t]{\textwidth}{\centering \scriptsize
			Personal use of this material is permitted. Permission from the author(s) and/or copyright holder(s), must be obtained for all other uses. Please contact us and provide details if you believe this document breaches copyrights.\\
			\noindent\makebox[\linewidth]{}
		}\hfil\hbox{}}%
	\def\@evenhead{\scriptsize\thepage \hfil \leftmark\mbox{}}%
	\def\@oddfoot{\parbox[t][\height][l]{\textwidth}{
			\vspace{-20pt}{\rule{\textwidth}{0.4pt}}\\ \footnotesize{\bf{\footnotesize\textcolor{red}{Hussein Naeem Hasas, "A Wearable Rehabilitation System to Assist Partially Hand Paralyzed Patients in Repetitive Exercises," In Proc. of the First International Scientific Conference Al-Ayen University, Nasiriyah, Thi-Qar, Iraq 2019.}}}\\
			\noindent\makebox[\linewidth]
		}\hfil\hbox{}}%
	\def\@evenfoot{\MYfooter}}

\makeatother
\begin{document}
\bstctlcite{IEEEexample:BSTcontrol}

\title{A Wearable Rehabilitation System to Assist Partially Hand Paralyzed Patients in Repetitive Exercises}

\author{\IEEEauthorblockN{Hussein Naeem Hasan}
\IEEEauthorblockA{\textit{Boimedical Engineering Department, College of Engineering} \\
\textit{University of Thi-Qar},
Thi-Qar, Iraq \\
husseinalshami@utq.edu.iq}
}

\maketitle

\begin{abstract}
The main purpose of the paper is development, implementation, and testing of a low cost portable system to assist partially paralyzed patients in their hand rehabilitation after strokes or some injures. Rehabilitation includes time consuming and repetitive exercises which are costly and demotivating as well as the requirements of clinic attending and direct supervision of physiotherapist. In this work, the system consists of a graphical user interface (GUI) on a smartphone screen to instruct and motivate the patients to do their exercises by themselves. Through the GUI, the patients are instructed to do a sequence of exercises step by step, and the system measures the electrical activities (electromyographic signals EMG) of the user’s forearm muscles by Myo armband. Depending on database, the system can tell whether the patients have done correct movements or not. If a correct movement is detected, the system will inform the user through the GUI and move to the next exercise. For preliminary results, the system was extensively tested on a healthy person.
\end{abstract}


\section{Introduction}
Brain stroke causes many impairments and physical disabilities for the survived persons. Hand impairment is one of the physical disabilities after stroke which in many cases can be healed totally or partially by physical therapy and rehabilitation. Traditional rehabilitation requires the patients to attend a physical therapy clinic and a direct interaction of an occupational therapist to do long lasting sets of repetitive exercises, so it will be costly and accompanied with patients discomfort and demotivating. In recent decade, researchers have investigated many feasible solutions for such problems. Many researchers exploited the advance in robotics to design a wearable robotic exoskeleton to help severely impaired patients in after stroke rehabilitation and doing their daily life activities \cite{jarrasse2014robotic, mulas2005emg, ho2011emg, stein2007electromyography, bae2012wearable, 4651199, 4303112}.
The exoskeleton consists of set of links driven by actuators to perform predefined movements or predicted movements that depend on sensors feedback. Ho \cite{ho2011emg} designed a robotic exoskeleton hand that is driven by electromyographic signals (EMG) to detect the patients’ intention to do hand opening and closing movements and help them to achieve the task. Mulas \cite{mulas2005emg} also designed an EMG controlled exoskeleton with graphical interface on a personal computer. The exoskeleton is actuated by motors to help patients with partial disability. 

In order to motivate the patients to do their exercises and get more self-involved in the rehabilitation process, researchers have exploited many approaches like augmented reality, virtual reality, computer games, and so on.  Alamri \cite{alamri2010ar} used the augmented reality approach to design a rehabilitation system called AR-REHAB. The later contains two subsystems for patient and therapist. The patient subsystem is responsible for giving exercises, creating virtual reality of the surrounding environment and monitoring the patient while the therapist subsystem gives evaluation of the patient performance and send the data to the assistant therapist. For more on the virtual reality, reader can find a good review in \cite{holden2005virtual}. In the computer games approach, Shusong \cite{5513218} developed a computer game controlled by wearable EMG sensors accompanied with a robotic toy to communicate with the patents. In this system, patient can perform exercises for the impaired muscles by playing the game with the help of the robotic toy commands. 

Many of the systems described above contain multiple costly hardware components as well as the requirements of personal computers and particular operation training. In this work, we keep in mind many factors to design our system like cost, simplicity, size, and portability. We get the benefit of the fact of common use of smartphones in many life's aspects and their simplicity and affordability to design our system. With advance in technology, Myo armband was designed by Thalmic Labs as wearable technology. The Myo armband contains eight EMG sensors to accurately measure the electrical activates of the arm muscles and recognize multiple gestures of the user hand and fingers. For the aforementioned reasons, our system consists only of a smartphone and a Myo armband. The system is designed to help partially impaired patients to do their rehabilitation exercises with simple graphical instructions and motivations. A graphical user interface (GUI) on the smartphone screen was developed to instruct the user to perform a particular exercise while the electromyographic signals of the user's muscles are collected by the Myo armband and sent to the smartphone. Depending on database, the user activities will be recognized and tell whether he/she has performed a correct exercise or not. The user will be informed through the GUI about his/her performance by emoji faces. If the user does the correct action, the system will prepare next exercise. 

The rest of the paper is organized as follows, in section \ref{rehab} the rehabilitation system and its main components will be described. Section \ref{method} describes the methodology of the system operation. Results will be discussed in section \ref{results}. And finally, section \ref{conc} for the conclusions and the future works.

\section{The Rehabilitation System \label{rehab}}
\begin{figure}[bp]
\centerline{\includegraphics[width=0.45\textwidth]{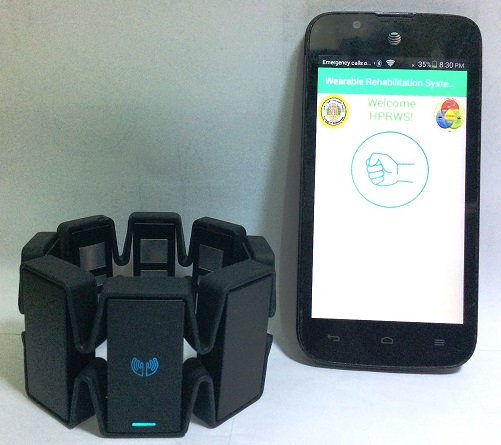}}
\caption{The rehabilitation system.}
\label{system}
\end{figure}

\begin{figure}[tph]
\centerline{\includegraphics[trim=1cm 16cm 2cm 1.5cm, width=0.65\textwidth]{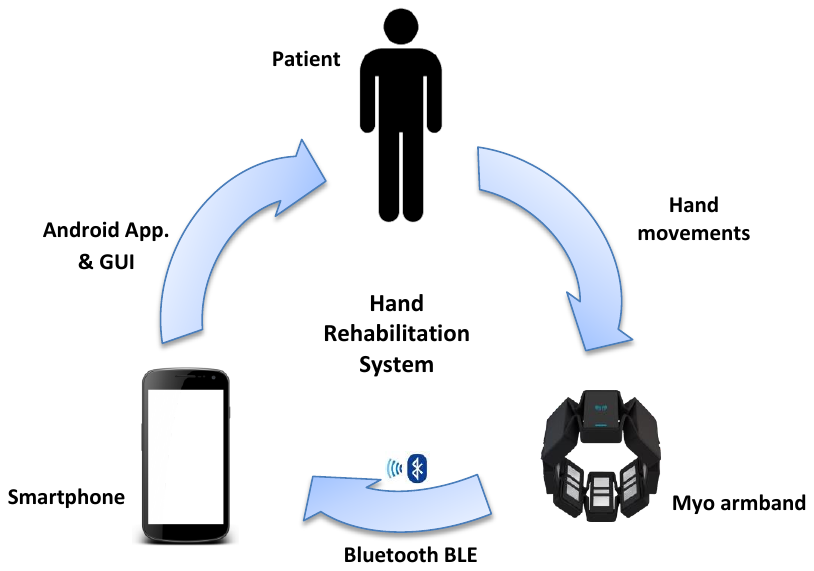}}
\caption{The schematic diagram of the rehabilitation system.}
\label{scheme}
\end{figure}

The system simply consists of two components, namely a Myo gestures control armband and a smartphone as shown in Fig.~\ref{system}. In addition, an application for the smartphone has been developed to handle the graphical user interface (GUI) as well as the the computational processing tasks and the wireless communication with the Myo armband through Bluetooth. Figure~\ref{scheme} shows a schematic diagram of the rehabilitation system. To explain the main functions in this project, detailed description will be provided for each component in the following subsections.
 
\subsection{The Myo Gestures Control Armband}
\begin{figure}[tp]
\centerline{\includegraphics[trim=1cm 13cm 2cm 1.5cm, width=0.5\textwidth]{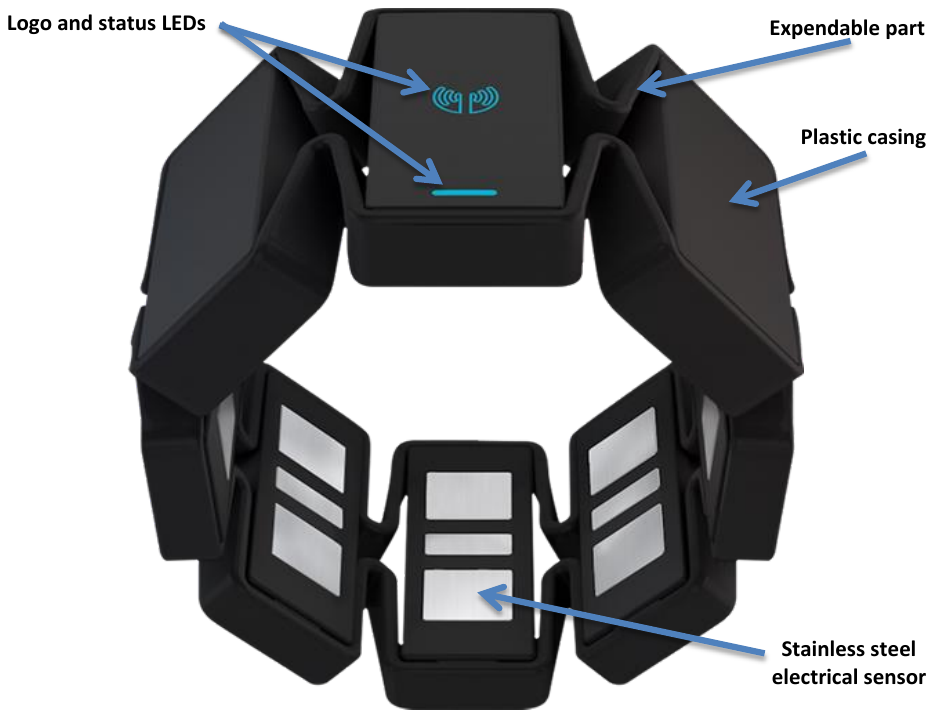}}
\caption{The main parts of the Myo armband, adapted from \cite{ThalmicPic}.}
\label{Myo}
\end{figure}

According to \cite{Thalmic}, Myo gestures control armband is a wearable technological device that is designed by Thalmic Labs for controlling purposes through hand gestures and movements. From the hardware aspect, Fig.~\ref{Myo} shows the main parts of the Myo armband. It consists of eight stainless steel surface mounted (sEMG) sensors to measure the electrical activities of the user' arm muscles. These eight bio-sensors arranged around the user's arm to detect any movement of fingers and wrist of the hand. In addition, there is (9DOF) Inertial Measurement Unit (IMU) that can detect and measure the hand motion in all directions and gives the motion data in the form of acceleration and orientation \cite{Abaid}. The IMU unit consists of 3D accelerometer, 3D gyroscope, and 3D magnetometer \cite{SATHIYANARAYANAN201550}.  More, there is an ARM Cortex M4 Processor to control all processing operations in the Myo armband system. The later can communicate wirelessly through Bluetooth 4.0 Low Energy (BLE). The entire system is contained in eight plastic casing and powered by a rechargeable lithium battery that can be recharged via USB Port. Moreover, The expendable parts are used to bind the plastic casings together and give flexibility for the armband to firmly fit the user's arm.

\begin{figure}[h!]
\centerline{\includegraphics[width=0.4\textwidth]{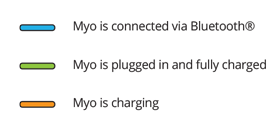}}
\caption{The status LED and the colors meaning, adapted from \cite{Thalmic}.}
\label{statusLED}
\end{figure}

From Fig.~\ref{Myo}, we can see two types of LEDs, namely the logo LED to indicate the synchronization state of the Myo armband while the status LED shows the charging and connection states through color coding as shown in Fig.~\ref{statusLED}

From the software aspect, Myo armband is compatible with Windows (7, 8, and 10) and Mac for computers while for tablets and smartphones, it can be easily used with Android and iOS. 
The principle function of the Myo armband is measuring the electrical activities of arm muscles of the patient during the exercises and process the data to detect whether the correct gesture has been done or not. Then, it sends the results to the smartphone.

\subsection{The Smartphone}
To keep the cost as low as possible, an android smartphone has been used in this project. Many android smartphones are relatively cheap and commonly used worldwide. Smartphone offers many useful accessible features that make it an promising and affordable choice for many engineering projects. It is a small sized, compatible platform with highly powered processing unit and sophisticated operational system. 

In addition, it contains a big enough touch screen that can be used for the graphical interface for the user in our project. Moreover, it contains an audio system which can be used to communicate and instruct the patient audibly as well as there is an internal and external memories for computational processing and storing the data. All of the aforementioned accessible features make the smartphone as an excellent choice for our project.

The main function of the smartphone in this project is explained through the description of the Android Application that has been developed to perform the following operations:
\begin{itemize}
\item Search and make connection with the Myo armband via Bluetooth.
\item Start the rehabilitation exercise and instruct the patient to follow the graphical and audio instructions.
\item Listen to the Myo armband in order to receive any available data.
\item Distinguish between correct or incorrect gestures and movements depending on database.
\item Display the results on the screen using patient friendly emoji, text, and audio voice in order to motivate him/ her to continue their exercise.
\item If the patient does the correct gesture or movement, the next exercise will be prepared.
\end{itemize}

\begin{figure}[htbp]
\centering
\includegraphics[trim=4.2cm 6.2cm 5cm 2cm, width=0.575\textwidth]{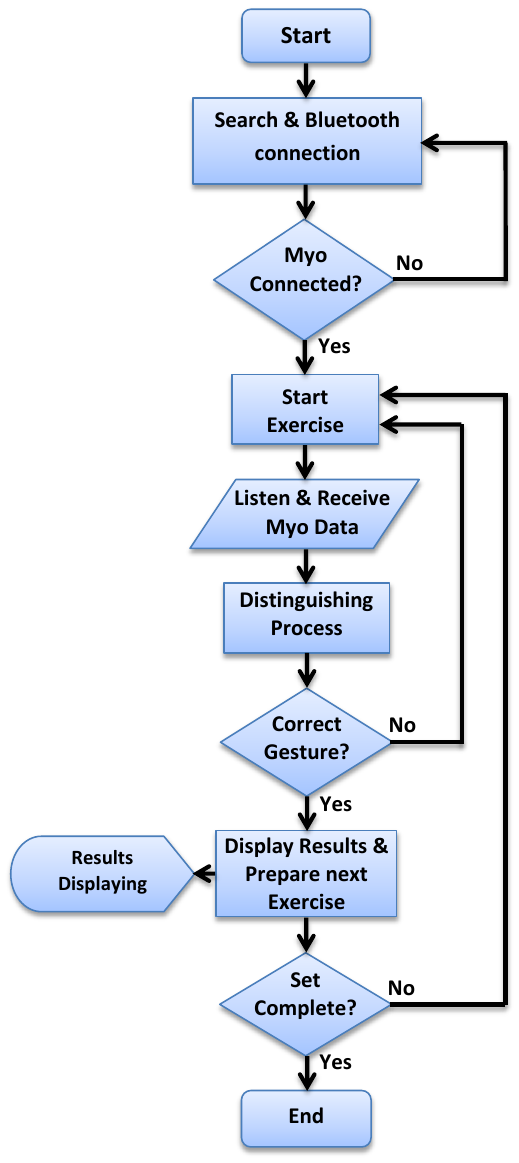}
\caption{Flow chart for the system's operations.}
\label{flowChart}
\end{figure}

Android Studio IDE from \textit{Google}\textsuperscript{\textcopyright} and Software Development Kit (SDK) for android from \textit{Thalmic Labs}\textsuperscript{TM} have been utilized to develop the Android application. A Huawei Android smartphone with android version (kitkat 4.4.3) has been used in our experiments. Figure \ref{flowChart} shows a flow chart for the system's operations.

\begin{figure}[bth]
    \centering
        \includegraphics[width=0.475\textwidth]{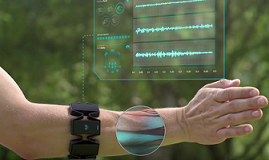}
    \caption{The "Wave out" gesture for synchronization, adapted from \cite{geekwire}.}
    \label{waveout}
\end{figure} 

\section{Methodology \label{method}}
The system is intended to help patient with partial paralysis to do some hand rehabilitation exercises after surviving from stroke. With this small-sized portable system, patients can do their exercises everywhere even when they are on the bed. The patient should wear the Myo armband on his/ her arm near to the elbow and do ”Wave out” gesture for synchronization as shown in Fig.~\ref{waveout}. A vibration feedback will be provided if the Myo armband syncs correctly. Then, open the android application to connect the Myo armband to the smartphone via Bluetooth. After connection, the system will start the exercise by displaying the finger and wrist movement and gesture on the screen and instruct the patient to do the displayed movement. The movement should be hold for five seconds and repeated in five or ten times in three sets or as instructed by the therapist. During the exercise all the muscles activities will be monitored and detected by the Myo armband and sent back to the smartphone. If the patient does the correct movement, he will be informed graphically by emoji faces, text, and voice records, otherwise the patient will be instructed to do the correct movement again as it is requested on the screen. 

\section{Results and Discussion \label{results}}
\begin{figure}[bth!]
    \centering
        \includegraphics[width=0.475\textwidth]{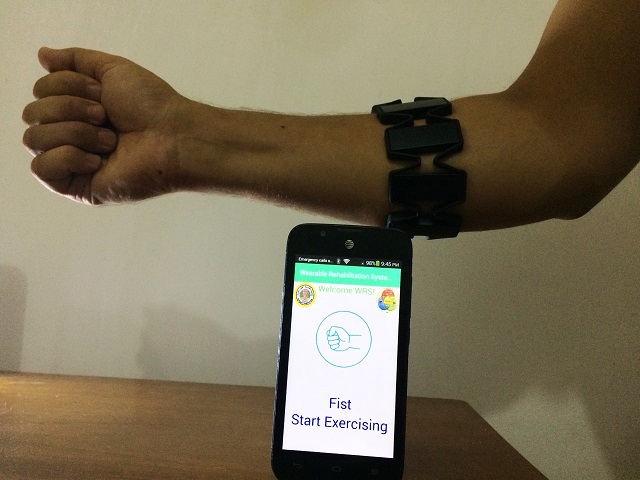}
    \caption{The start of exercise.}
    \label{Result1}
\end{figure}

For the purpose of this paper, preliminary results were obtained by testing the system on two healthy subjects (male and female with 33 and 26 years old respectively). Figure~\ref{Result1} shows the user’s hand with the Myo armband on his arm while there is a graphical instruction to do fist movement is displayed on the smartphone screen with text indicating the start of exercise. 

\begin{figure}[btp!]
    \centering
        \includegraphics[width=0.475\textwidth]{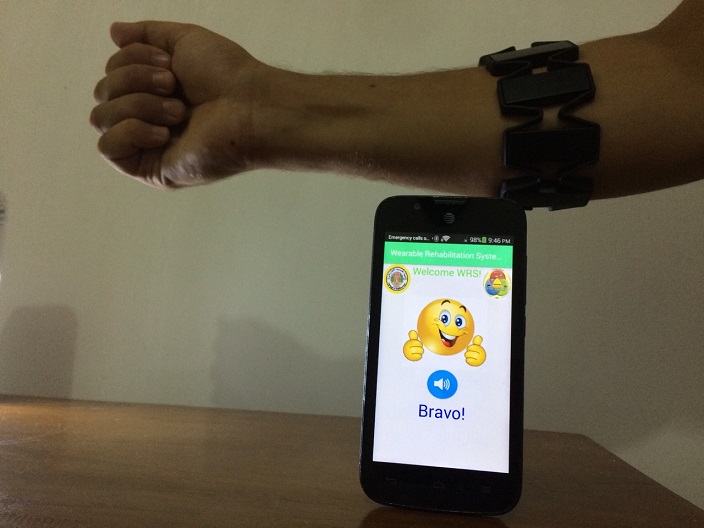}
    \caption{The correct movement result.}
    \label{Result2}
\end{figure}

\begin{figure}[bth!]
    \centering
        \includegraphics[width=0.475\textwidth]{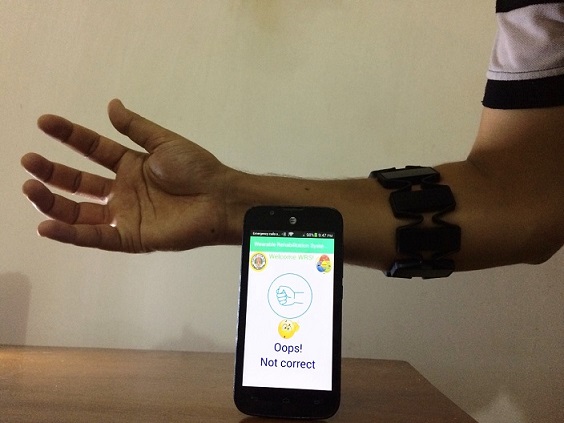}
    \caption{The incorrect movement result.}
    \label{Result3}
\end{figure}

\begin{figure}[bth!]
    \centering
        \includegraphics[width=0.475\textwidth]{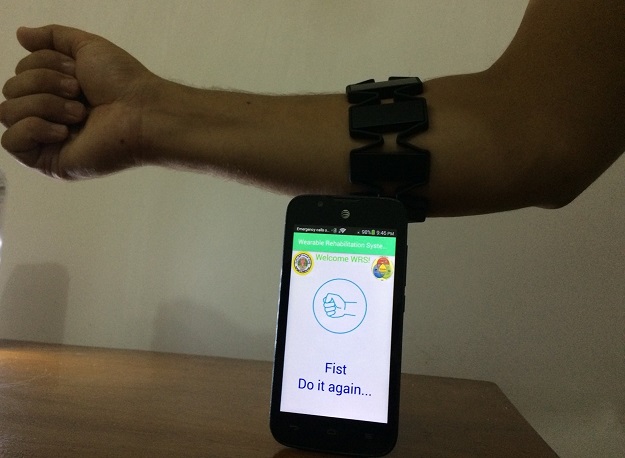}
    \caption{Waiting for the next movement in the same set (fist).}
    \label{Result4}
\end{figure}

When the user did the correct movement, the result was displayed on the screen as shown in Fig.~\ref{Result2}. We can see the happy emoji face that indicates positive progress with some text to motivate the user to continue. On the other hand, if the user did incorrect movement, we can see the graphics in Fig.~\ref{Result3} indicate that the movement was not correct. Figure~\ref{Result4} shows that the system is waiting for the user to do the (Fist) movement again to complete the exercise set.  

\begin{figure}[btp]
    \centering
        \includegraphics[width=0.475\textwidth]{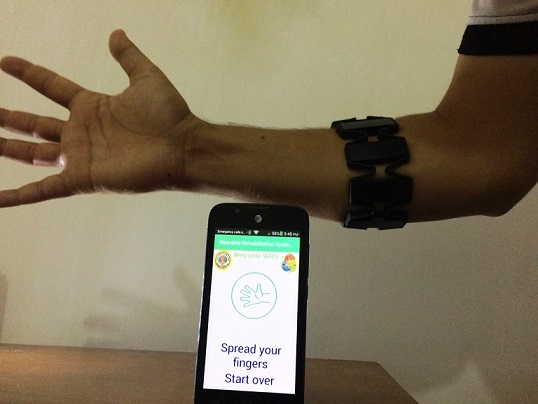}
    \caption{Fingers spreading movement.}
    \label{Result5}
\end{figure}

\begin{figure}[tb!]
    \centering
        \includegraphics[width=0.475\textwidth]{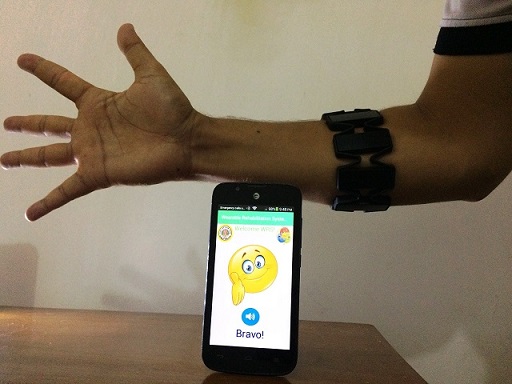}
    \caption{The correct movement result.}
    \label{Result6}
\end{figure}

\begin{figure}[bth!]
    \centering
        \includegraphics[width=0.475\textwidth]{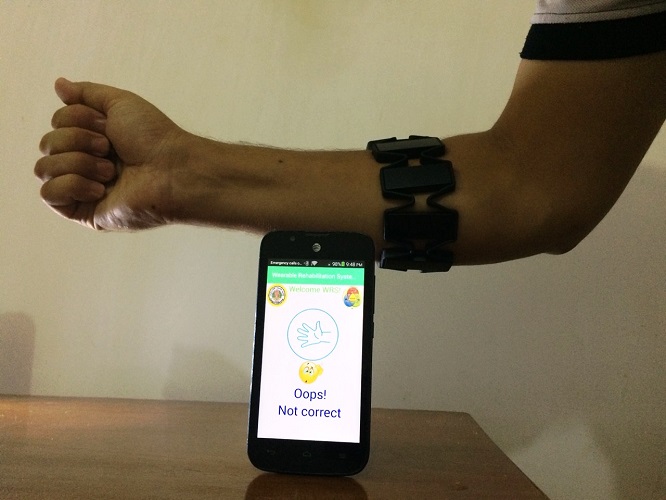}
    \caption{The incorrect movement result.}
    \label{Result7}
\end{figure}

\begin{figure}[bth!]
    \centering
        \includegraphics[width=0.475\textwidth]{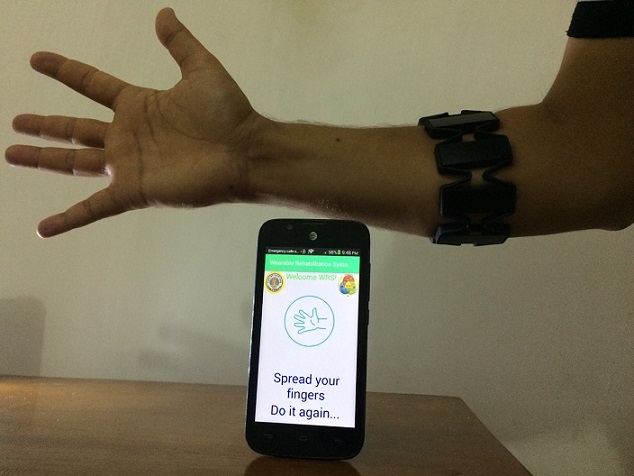}
    \caption{Waiting for the next movement in the same set(fingers spreading).}
    \label{Result8}
\end{figure}

After the first movement exercise set was done correctly, the next exercise for different movement was prepared as shown in Fig.~\ref{Result5}. Figures~\ref{Result6} to \ref{Result8} show the rest results of the second exercise set that happened in the same sequence in the first exercise set. We show only the first and second set of exercises to explain the preliminary results and effectiveness of the system while there are multiple sets of exercises.

\section{Conclusion and Future Works \label{conc}}
Recovery from disability after stroke in curable cases requires starting rehabilitation exercises as soon as possible to obtain good results. In this work, a wearable rehabilitation system based on smartphone technology and Myo armband was designed, implemented, and tested to help partially disabled patient to do their exercises by their own. The preliminary results showed a promising approach that can be further developed to produce more sophisticated system. For the future work, we will continue working on the system to develop a platform that allows the therapist to monitor the patient's performance through internet connection or by reviewing the stored performance data. In addition, we will cooperate with caregiver clinics to test the final version of the system on patients.

\bibliographystyle{IEEEtran}
\bibliography{reference}
\end{document}